\documentclass[12pt]{article}
\usepackage{times}

\usepackage[utf8]{inputenc}
\usepackage{authblk}
\usepackage{graphicx}
\usepackage{siunitx}
\usepackage{etoolbox}
\usepackage[para]{footmisc}

\DeclareUnicodeCharacter{2060}{}
\DeclareUnicodeCharacter{0301}{}

\usepackage{cite}

\begin{document}

\title{Aligned Stacking of Nanopatterned 2D materials -- Towards 3D printing at atomic resolution}

\author[1,2,\thanks{jonas.haas@uni-tuebingen.de}]{Jonas Haas}
\author[1,2]{Finn Ulrich}
\author[1,2]{Christoph Hofer}
\author[3]{Xiao Wang}
\author[4]{Kai Braun}
\author[1,2,\thanks{jannik.meyer@uni-tuebingen.de}]{Jannik C. Meyer}
\date{}

\renewcommand\Affilfont{\fontsize{9}{10.8}\itshape}

\affil[1]{Institute of Applied Physics, Eberhard Karls University of Tuebingen, Auf der Morgenstelle 10, D-72076, Tuebingen, Germany}
\affil[2]{Natural and Medical Sciences Institute at the University of Tuebingen, Markwiesenstr. 55, D-72770 Reutlingen, Germany}
\affil[3]{School of Physics and Electronics, Hunan University, Changsha, Hunan 410082, China}
\affil[4]{Institute of Physical and Theoretical Chemistry, Eberhard Karls University of Tuebingen, Auf der Morgenstelle 18, D-72076, Tuebingen, Germany}

\maketitle

\begin{abstract}
Two-dimensional materials can be combined by placing individual layers on top of each other, so that they are bound only by their van der Waals interaction. The sequence of layers can be chosen arbitrarily, enabling an essentially atomic-level control of the material and thereby a wide choice of properties along one dimension. However, simultaneous control over the structure in the in-plane directions is so far still rather limited. Here, we combine spatially controlled modifications of 2D materials, using focused electron irradiation or electron beam induced etching, with the layer-by-layer assembly of van der Waals heterostructures.  A novel assembly process makes it possible to structure each layer with an arbitrary pattern prior to the assembly into the heterostructure. Moreover, it enables a stacking of the layers with accurate lateral alignment, with an accuracy of currently 10nm, under observation in an electron microscope. Together, this enables the fabrication of almost arbitrary 3D structures with highest spatial resolution.
\end{abstract}

\vspace{10pt}\noindent\textbf{Keywords:} 2D materials, van der Waals heterostructures, nanopatterning, electron microscopy, nanoscale assembly, 3D printing

\section{Introduction}

Structuring matter on the smallest dimensions is a key capability for many areas of technology, most prominently for information processing but also for sensing, nanofluidics, optics, catalysis, medicine and many more \cite{Bhushan2007,Luttge2011,Kandelousi2018,Bocquet2020,Tibbals2017,Heath2015}. As an observation that has become known as Moore's law \cite{Schaller1997,MacK2011}, the feature sizes in information processing devices have continuously shrunk over many decades, now approaching limitations given by the granularity of the atomic structure. However, current computing devices are essentially planar. Methods for generating arbitrary 3D structures are commonly referred to as "3D printing". Also in this area, the minimum feature sizes that can be defined are continuously shrinking as new approaches are invented and developed, although the 3D resolution is still larger than what can be achieved in planar structures \cite{Engstrom2014,Farahani2016, Hirt2017, Lin2020,Liashenko2020,Jesse2016}. Currently, optical methods can be pushed to a resolution of about 100nm \cite{Lin2020}. Electron- or ion-beam induced deposition can reach a resolution in the nanometer-range for planar structures on thin membranes \cite{VanDorp2005}, but generally the resolution is limited by the interaction volume rather than the beam diameter, which also depends on the target geometry, and for 3D structures is typically in the range of tens of nanometers \cite{Hirt2017}. Another important aspect for 3D printing techniques is the (often rather limited) choice of the materials that can be integrated with each other for fabricating functional devices \cite{Deore2021}.

Two-dimensional materials (2D materials) are isolated, one unit cell (or half a unit cell) thick layers, most of which can be described as an individual plane extracted from a van der Waals bonded layered material \cite{Novoselov2005,Mas-Balleste2011,Das2015,Novoselov2016,Zeng2018}. Some of them are only one atom thick (e.g. graphene or single-layer hexagonal boron nitride), while many more examples can be given for structures with two, three or more atomic layers. Importantly, these layers can be assembled in any arbitrary sequence, yielding so-called van der Waals (vdW) heterostructures \cite{Ponomarenko2011,Britnell2012,Haigh2012a,Geim2013,Novoselov2016,Kang2017,Frisenda2018,Zeng2018,Ma2019,Hemnani2019,Gant2020,Guo2021,Quellmalz2021}. This allows combining unique properties of the constituents, and may also lead to entirely new features due to the interaction between the layers. For example, hexagonal boron nitride (hBN) is not only an excellent substrate for graphene \cite{Dean2010b}⁠, but can also have a profound influence on graphene’s electronic properties that sensitively depends on the relative alignment \cite{Kan2012,Hunt2013} of the sheets. The assembly of conducting (e.g. graphene), insulating (e.g. hBN) and semiconducting layers (e.g. molybdenum disulfide, MoS$_{2}$) has led to a wide range of new device geometries such as transistors with ultra-thin gate dielectric \cite{Lee2013b}⁠, tunneling transistors \cite{Britnell2012}⁠, memory devices \cite{SupChoi2013}⁠, oscillators \cite{Mishchenko2014}⁠, photo-transistors\cite{Roy2013}⁠ and photo-voltaic devices \cite{Britnell2013}. Modifications of the electronic structure often rely on Moiré effects or other naturally occurring short-range periodicities \cite{Celis2018}, with a prominent example being the superconductivity observed in graphene bilayers with a magical misorientation angle of 1.1° \cite{Cao2018}⁠. Moiré lattice potentials provide an easy access to periodic variations in the few-nm range, however, they can not be directly used to fabricate more complex potential landscapes.  Besides electronic and optical applications, vdW heterostructures are important for molecular sieves \cite{Abraham2017}, nanopores \cite{Zou2020}, catalysis \cite{Ren2019}, or nanoscale liquid cells \cite{Kelly2018}, just to name a few. 

2D materials can be structured by conventional lithographic techniques, after they are deposited on a substrate \cite{Stampfer2008, Jessen2019}. However, much higher resolution can be achieved if 2D materials are available as free-standing membranes: In particular, the direct cutting of graphene \cite{Fischbein2008,Song2011,Borrnert2012,Xu2013}, other 2D materials \cite{Liu2014,MasihDas2016}, or complete vdW heterostructures \cite{Clark2019a,MasihDas2020} by the focused electron beam in a transmission electron microscope (TEM) or scanning transmission electron microscope (STEM) makes it possible to define feature sizes in the nanometer range.  Similarly, electron-beam induced etching (EBIE) \cite{Spinney2009, Sommer2015} or cutting of 2D materials by ion beams \cite{Bell2009,Fox2015} is preferentially done with free-standing membranes. Individual impurity atoms in graphene membranes can even be relocated to specific atomic sites \cite{Susi2017d,Dyck2017,Tripathi2018}, enabling a true atomic-level control, comparable to the relocation of atoms on surfaces using scanning tunneling microscopy \cite{crommie_confinement_1993}.  However, again these structures are limited to two dimensions, and the integration with other materials, e.g. additional nanostructured layers, is not straightforward.

\begin{figure}
\centering
\includegraphics[width =\linewidth]{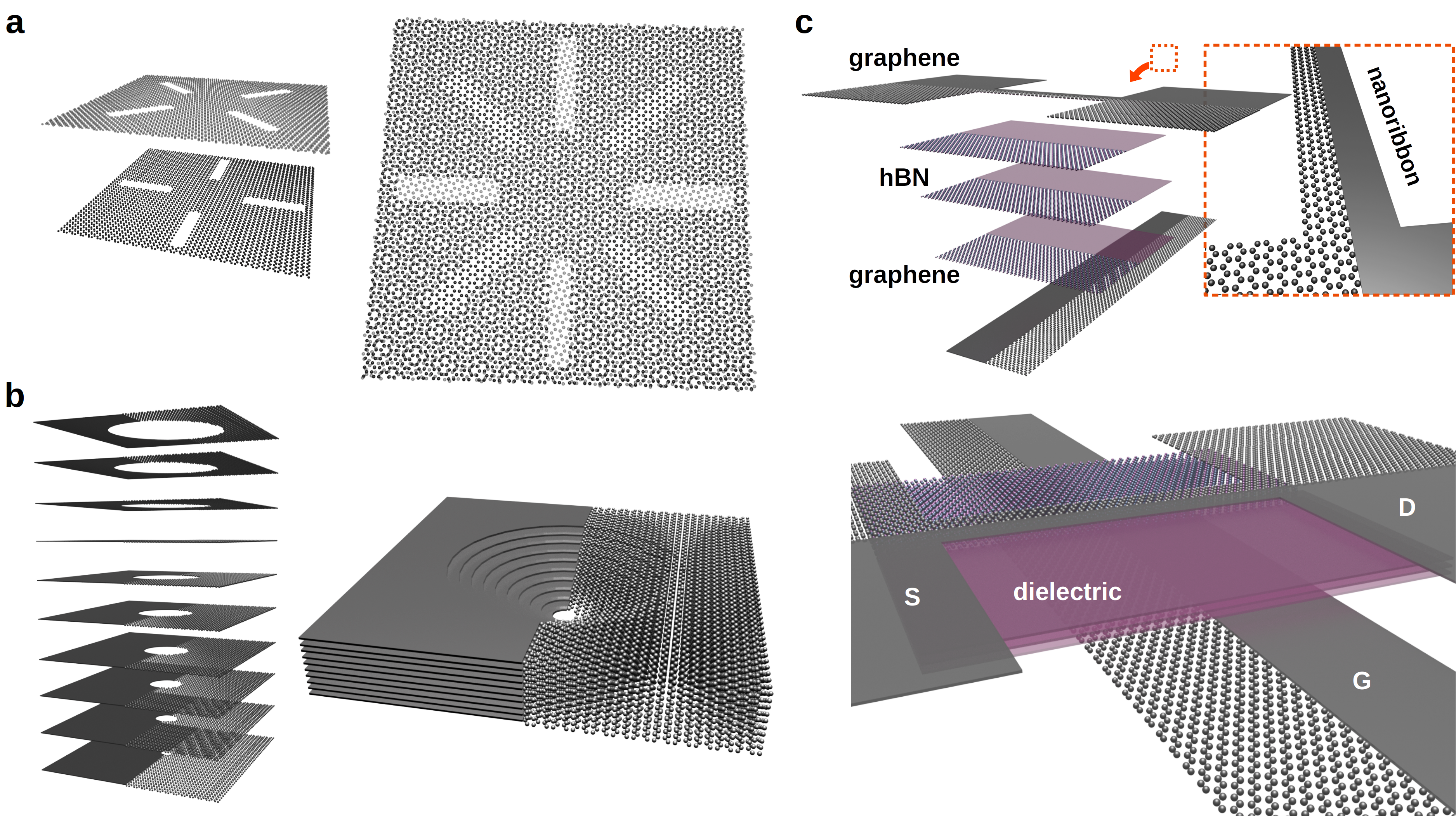}
\caption{Formation of arbitrary 3D structures by the assembly of pre-patterned 2D material layers. (a+b) Illustration of test structures shown in this work, namely (a) two rotated crosses and (b) a nanoscale funnel built from holes of increasing size. (c) Hypothetical transistor structure consisting of graphene leads, a graphene nanoribbon as active element, hBN gate dielectric and a graphene gate electrode.}

\label{fig:3dprint}
\end{figure}

In this work, we developed an approach that combines the almost-atomic resolution structure definition in free-standing 2D material membranes with the assembly of pre-structured layers into van der Waals heterostructures. Fig. \ref{fig:3dprint} schematically illustrates the idea: A structure is defined in every one- or few-atom thin layer, and the layers are placed on top of each other with a precise lateral alignment. The concept can be categorized as 3D printing, which (in most of its implementations) also proceeds by forming a target structure layer by layer. Critically, the lateral alignment between the layers must be of a similar accuracy than the targeted resolution. Here, we repeatedly achieve an accuracy of 10nm, although this is probably not a fundamental limitation of the approach. We also point out that the material for each layer can in principle be chosen from the vast repertoire of available 2D materials, which should make it possible to directly create functional devices. 

\section{Experiments and Results}

\subsection{General considerations}

\begin{figure}
\includegraphics[width=1.0\textwidth]{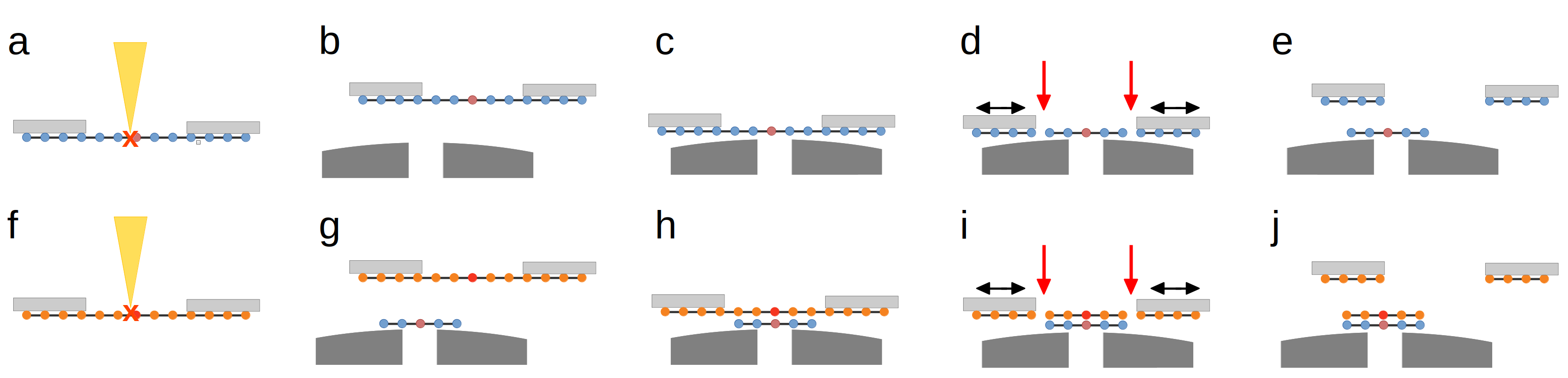}
\centering \caption{Schematic of the stacking process. (a) The 2D material, which spans a hole in the support frame (e.g. TEM grid), is patterned by focused electron irradiation (modified area indicated as red dot). (b+c) The membrane is moved (under SEM observation) onto a second support with a smaller hole. A slight curvature of the second support frame makes it possible to bring the two surfaces into contact at the desired point. (d) Lateral motion of the support frames relative to each other (black arrows) leads to a rupture of the membrane around the edges of the larger hole (red arrows). (e) Upon separation of the support frames, part of the 2D material stays on the target frame. (f-j) The process is repeated with additional layers, resulting in a stack of 2D material layers.}

\label{schematic} 
\end{figure}

To enable structured modifications by focused electron irradiation at highest resolution, it is essential that the individual layers are free-standing at this step, which is prior to the stacking.  Similarly, it is important that the aligned stacking is done with visual feedback from an electron microscope, hence it has to be done in a vacuum (using nano-manipulators). These requirements are in conflict with previously published methods for the deterministic transfer of 2D materials and their assembly into heterostructures
\cite{Kang2017,Frisenda2018,Ma2019,Hemnani2019,Gant2020,Guo2021,Quellmalz2021}, all of which employ a transfer medium in contact with the layers, and in some cases a chemical removal of the transfer layer. The concept of our layer transfer is schematically shown in Fig. \ref{schematic}: We start with a 2D material layer as free-standing membrane, attached to a relatively large support frame (10µm diameter in our case). Into this membrane,  a pattern is written using focused electron irradiation in the (S)TEM, or by EBIE in the scanning electron microscope (SEM). The membrane is brought into contact with a second, smaller support frame. The surface around this second hole is curved, to facilitate a contact at the desired area in spite of unavoidable small tilt between the two frames. Upon contact, the 2D material membrane sticks to the surface of the smaller support frame, and spans the hole in this frame.  Importantly, this adhesion is strong enough that the material can be detached from the initial frame: By moving the initial support frame slightly sideways (black arrows in Fig. \ref{schematic}d), the material breaks off from this frame around the edges (red arrows in Fig. \ref{schematic}d). Now, the process can be repeated with the next layer (Fig. \ref{schematic}f-j).

\subsection{Implementation}

\begin{figure}
    \centering
    \includegraphics[width = .6\linewidth]{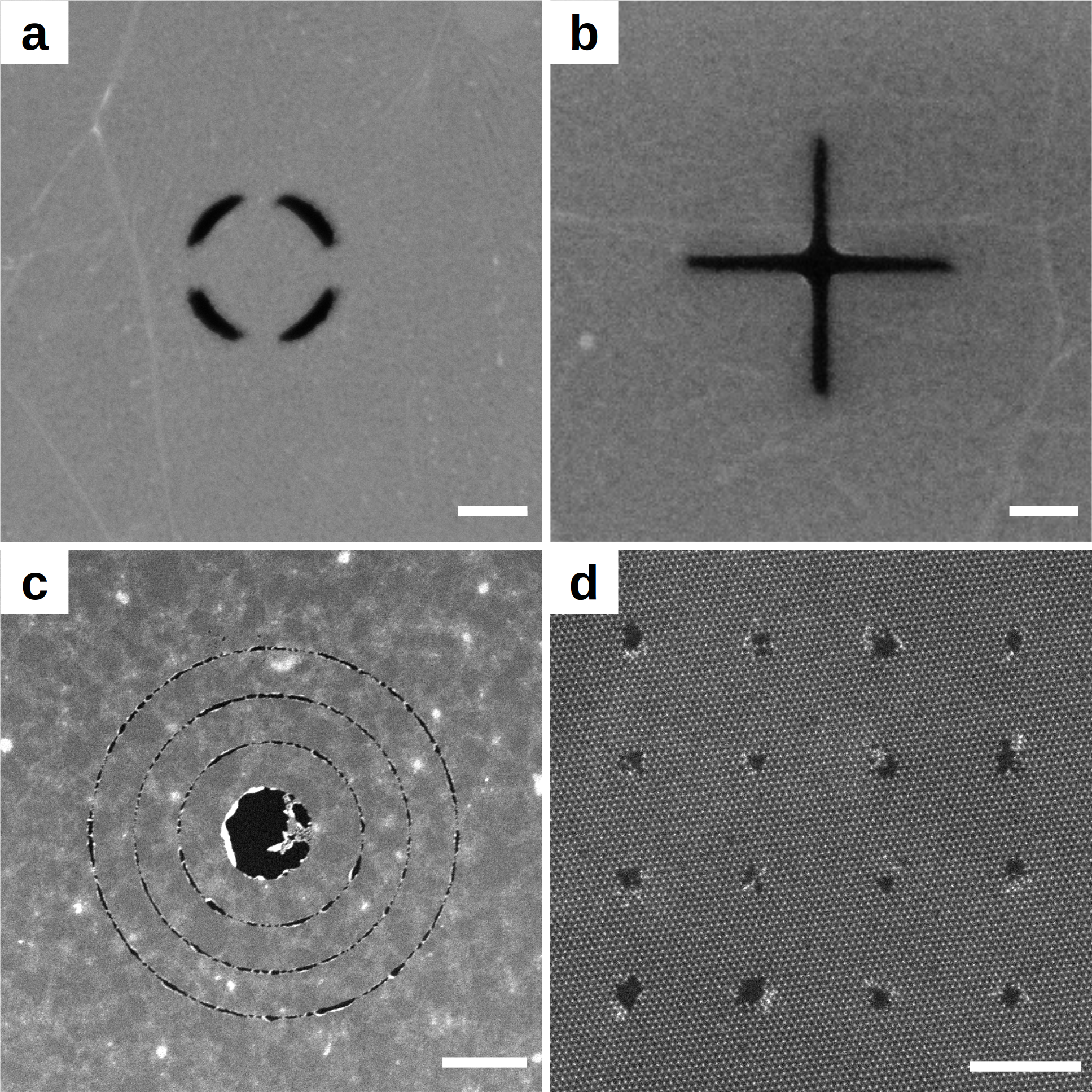}
    \caption{Examples of patterning the free-standing membranes on the initial support frame before stacking. (a+b) H$_{2}$O assisted EBIE is used to pattern large features into graphene. (a) Four circular segments and (b) a cross are introduced. (c+d) More filigree patterns can be created using STEM. (c) The image shows a hole with concentric rings patterned into a MoS$_{2}$ membrane. (d) A regular array of \SI{0.5}{} to \SI{1}{\nano\meter} sized holes is introduced into WS$_{2}$ by means of an aberration-corrected STEM which allows manipulation at a true atomic-level. Scale bars are (a+b) \SI{500}{\nano\meter}, (c) \SI{50}{\nano\meter} and (d) \SI{5}{\nano\meter}.}
    \label{fig:patterning}
\end{figure}

After introducing the concept, we now show the practical implementation. So far, we have prepared stacks of patterned Tungsten disulfide (WS$_{2}$), Molybdenum disulfide (MoS$_{2}$), and graphene. These 2D materials are prepared into free-standing membranes, and patterns are written into the membranes. For writing patterns into the WS$_{2}$ or MoS$_{2}$, we used a 200kV electron beam in a (S)TEM.
For graphene, we alternatively use EBIE for removing material in a controlled pattern. This is done using a 3kV electron beam in a scanning electron microscope in presence of water vapor (see methods for details). While EBIE has less spatial resolution than sputtering at high voltages, it requires less time for structuring larger areas. Moreover, it simplifies the process as it can be done in the same device as the stacking.
Fig~\ref{fig:patterning} shows exemplary free-standing membranes patterned by EBIE and STEM, respectively.

\begin{figure}
\includegraphics[width=1.0\linewidth]{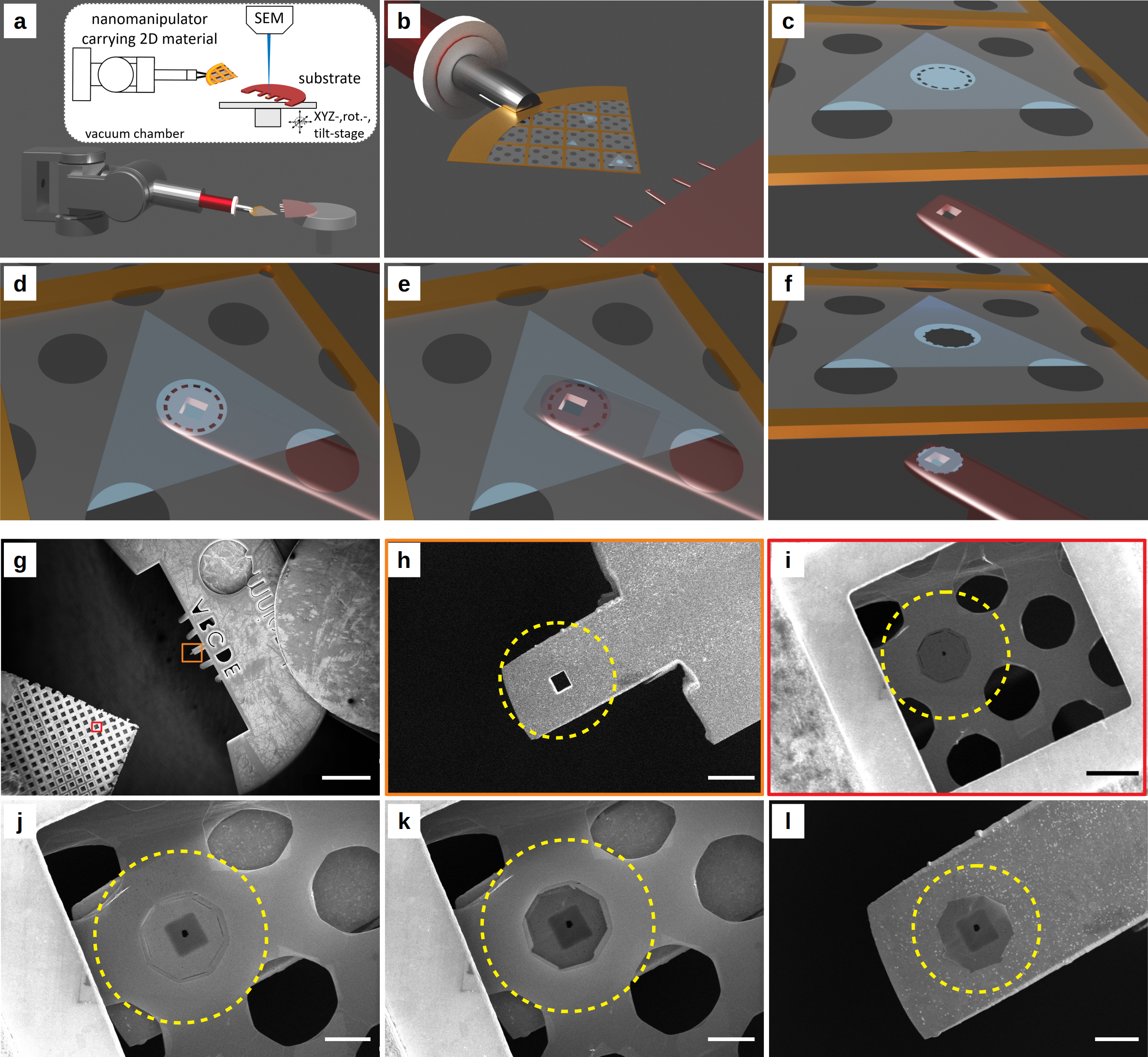}
\centering \caption{Controlled placement of a 2D material layer. (a-f) Schematic illustration of the procedure, and (g-l) SEM images of the process (please see also the supplementary video). (a+b+g) A quarter of a TEM grid carrying the 2D material is clamped into a nanomanipulator (left). The target substrate, one finger of a FIB lift-out grid (b+g right, c bottom) that was further cut into the desired shape by FIB, is mounted on the six-axis SEM stage. (c-e, i-k) One of the 10µm diameter holes in the grid, spanned by the structured 2D material, is located above the hole in the target frame and brought into contact. (f+l) The membrane breaks off from the TEM grid and remains on the target substrate. Scale bars are (g) \SI{500}{\micro\meter}, (h+i) \SI{10}{\micro\meter} and (j-l) \SI{5}{\micro\meter}.}
\label{fig:StackingSequence}
\end{figure}

Now we turn to the most critical aspect of placing and stacking the 2D materials with precise lateral alignment. 
An illustration of the experimental setup with the manipulation stages, which is used to carry out the presented process (except for patterning individual layers) is shown in  Fig. \ref{fig:StackingSequence}a. The individual layers are attached to a TEM grid, cut into one quarter. The target substrate is made from one of the fingers in a FIB lift-out grid, by making it more narrow, making it thinner, and inserting a hole. This is done by FIB, prior to placing the 2D materials, as shown in the supplementary information. For accurate alignment, the suspended, pre-structured layers on the TEM grid are held in a nanomanipulator which is attached to the SEM chamber. The target substrate is mounted on the six-axis SEM stage.

A schematic image sequence of the process is shown in Fig. \ref{fig:StackingSequence}a-f. 
First, the grid is brought above the target substrate, such that the free-standing membrane is centered on the hole in the target (see Fig. \ref{fig:StackingSequence}b-d). Now, the target substrate is moved upwards until a contact is made. The contact can be noticed by a clear change in contrast as illustrated in Fig. \ref{fig:StackingSequence}e. Slight lateral motion of the TEM grid leads to a rupture of the 2D material at the edge of its connection to the TEM grid. Breaking points, as shown in Fig. \ref{fig:StackingSequence}c-f as circular cutouts, can be introduced (already during patterning) to promote tearing-off. Now, the TEM grid can be moved upwards, leaving the membrane placed into the desired specific location on the target substrate (Fig. \ref{fig:StackingSequence}f). 

Fig. \ref{fig:StackingSequence}g-l shows an SEM image sequence of placing one graphene membrane. 
A close-up of the target substrate is shown in Fig. \ref{fig:StackingSequence}h, while Fig. \ref{fig:StackingSequence}i shows the membrane into which a hole was cut by EBIE suspended on the TEM grid. The grid is brought close to the target substrate, such that the hole in the graphene is centered on the hole in the target. As described above, the membrane is placed by bringing the membrane and the target in contact, notably by a change in the SEM image contrast (see Fig. \ref{fig:StackingSequence}j vs. Fig. \ref{fig:StackingSequence}k and video in supplement). Now, the membrane is torn off by slight lateral motions. Importantly, this motion does not cause any displacement of the membrane with respect to the substrate as their adhesion is strong enough, but only rupturing from the TEM grid at the predetermined breaking points. Fig. \ref{fig:StackingSequence}l shows the graphene layer after placement on the target substrate. This process is repeated for placing additional layers onto the first one. A video showing the aligned stacking of two patterned graphene layers is given in the supplement.

\begin{figure}
    \centering
    \includegraphics[width = 1\linewidth]{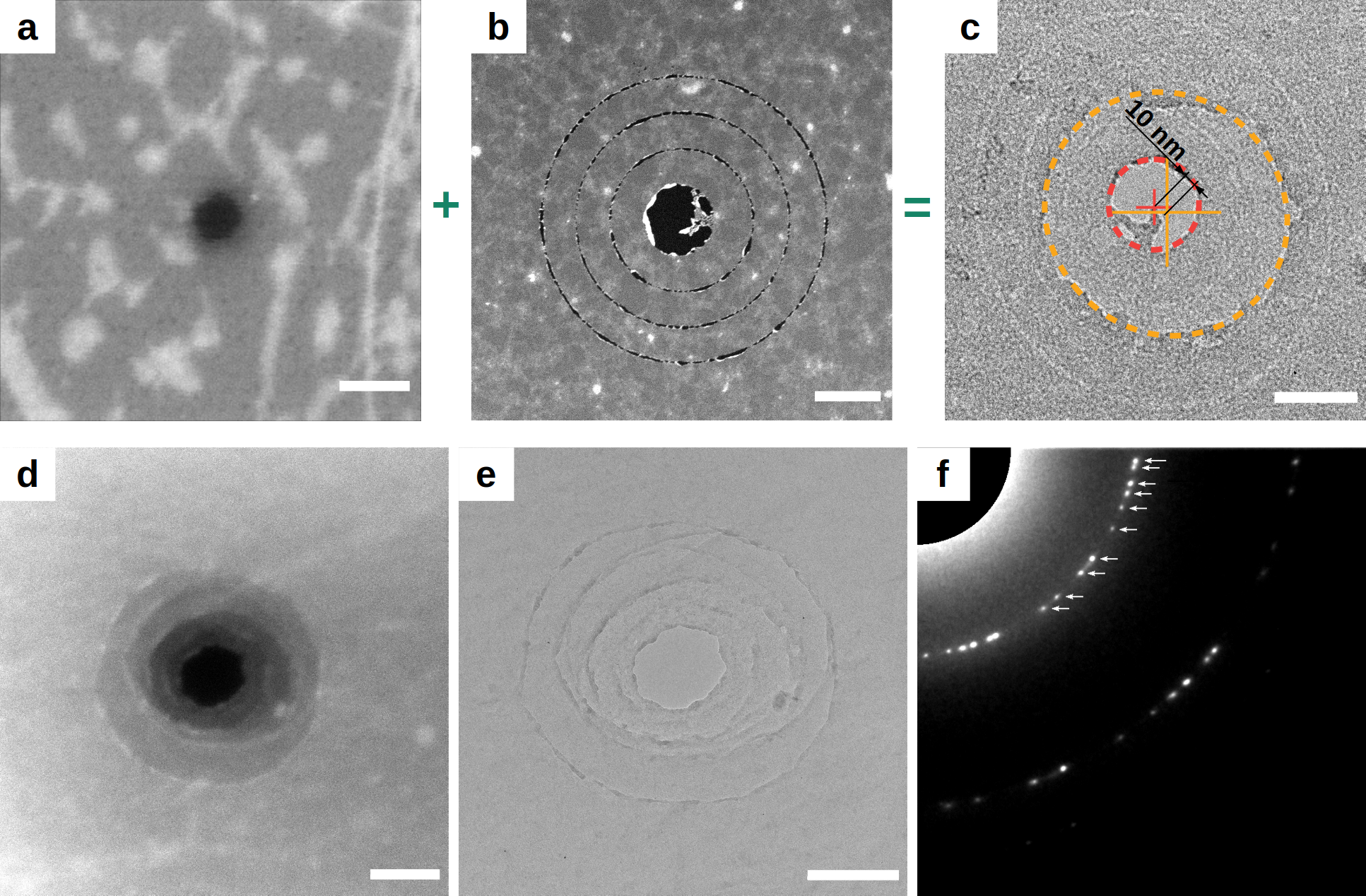}
    \caption{Examples of the layer-by-layer assembly. (a-c) Assembly of patterned graphene and MoS$_{2}$ membrane. (a) Hole in graphene, made by EBIE. (b) Central hole and concentric circles in MoS$_{2}$, made by sputtering in the STEM. (c) TEM image after aligned placement of the two structures. For illustration, the hole in graphene and in MoS$_{2}$ are marked in orange and red, respectively. The offset (distance between the centers of the holes) in this case is 10nm. 
    (d-f) Assembly of ten graphene layers forming a funnel. Each layer was patterned by EBIE before stacking with holes of different sizes and then stacked in sequence of increasing hole size. (c) shows the SEM and (d) TEM image of the assembly. In the diffraction pattern (f), the arrows indicate the ten different layers. SEM images after each assembly step with the respective offset is given in Fig. S2 in the supporting information. 
    Scale bar is (a) \SI{200}{\nano\meter}, (b+c) \SI{50}{\nano\meter} and (d+e) \SI{200}{\nano\meter}.}
    \label{fig:stackexamples1}
\end{figure}

\begin{figure}
    \centering
    \includegraphics[width = 1\linewidth]{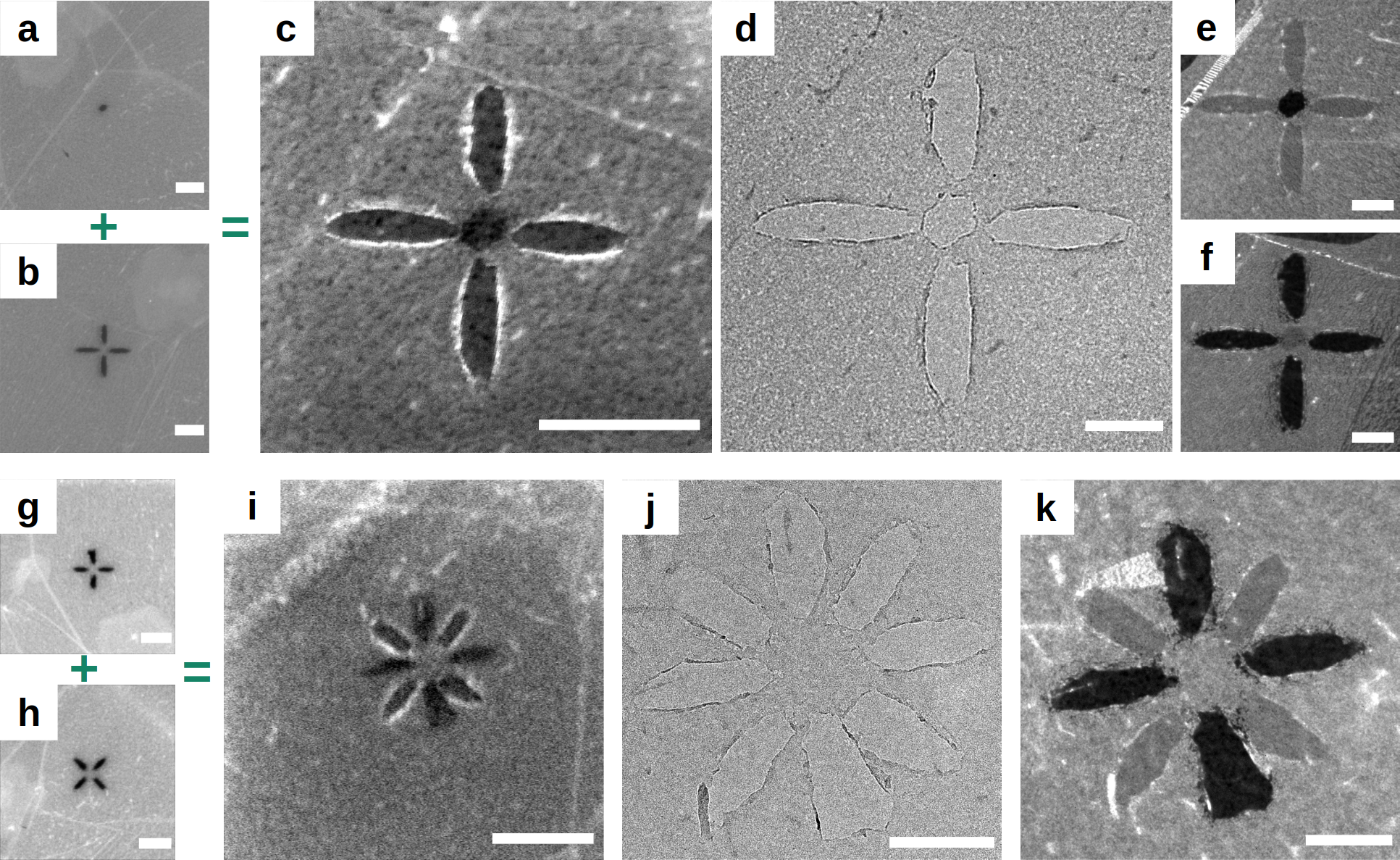}
    \caption{Examples of the layer-by-layer assembly.
    (a-f) Crosshair built from two graphene layers patterned by EBIE in the SEM with (a) a hole and (b) a cross. (c) SEM and (d) TEM image of the assembly. (e) and (f) are darkfield TEM images highlighting the different lattices of the respective layers.
    (g-k) Assembly of two crosses patterned in graphene by EBIE. (g) and (h) showing SEM images after patterning the membranes. The resulting stack is shown in (i-k) as SEM, TEM and darkfield TEM image (highlighting one of the two layers), respectively. The \SI{45}{\degree} relative angle between the two crosses as defined in the individual layers (g+h) is also maintained upon stacking (i-k). 
    Scale bar is (a-c, g-i) \SI{500}{\nano\meter}, (d-f+j+k) \SI{200}{\nano\meter}.
    }
    \label{fig:stackexamples2}
\end{figure}

Since the 2D materials are semi-transparent in SEM, features in both layers can be observed during placement of a layer, and hence an accurate alignment using the nanomanipulator is possible. It is worth noting that this applies both to the exact alignment of the membranes to each other as well as of the membranes on the substrate itself. While the layers are still at a distance, they can also be distinguished by the difference in focus height of the SEM (see supplementary video). For rough alignment, the secondary electron (SE) or the Inlens-SE detector of the SEM can be used. The combination of the Inlens-SE detector with a STEM detector, however, facilitates the observation of features in the already deposited layers for an accurate alignment. Figure \ref{fig:stackexamples1} and Figure \ref{fig:stackexamples2} show SEM and TEM images of layer structures prior to stacking and after stacking. 

In order to assess the accuracy of the lateral alignment, we analyze the offset between the patterns of the membranes in the final assembled structure. For this purpose, we measure the distance between the centers of the hole patterns, as exemplified in Figure~\ref{fig:stackexamples1}c. Here, the offset between the hole in graphene and the hole in MoS$_{2}$ is displaced by about \SI{10}{\nano\meter}. For the assembly of the ten graphene layers (see Figure \ref{fig:stackexamples1} d-f), the average offset was \SI{17}{\nano\meter}, with a minimum of \SI{7}{\nano\meter} and a maximum of \SI{35}{\nano\meter} (see also Figure S2 in the supplement).

\subsection{Integrating non-contiguous layer structures}

For many types of 3D printing, a limitation of the possible structures arises from the requirement that the structures must be stable at any time during the fabrication, i.e., already when the structure is only partly finished.  In the approach as described so far, even every layer must be stable by itself (as it is separately transferred), which means it can not contain isolated blocks of the material.

\begin{figure}
    \centering
    \includegraphics[width = .5\linewidth]{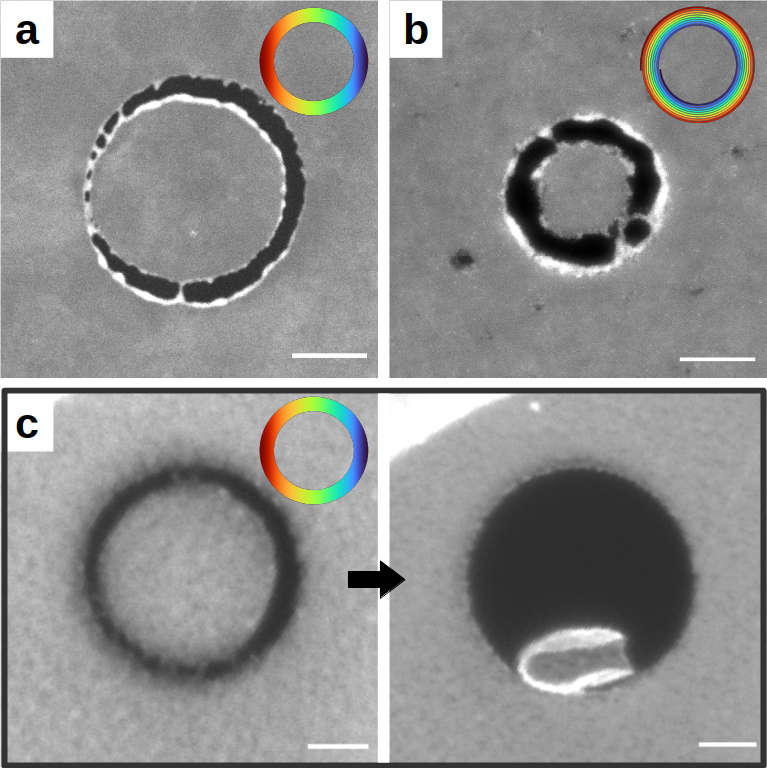}
    \caption{Examples for patterning "nearly" free-standing discs. (a+b) TMD membranes are introduced by sputtering in STEM. The sequence of writing is illustrated by the colored circles with writing from blue to red. (a) The circle is written column by column from right to left resulting in a shifted disc. In (b), the circle is written spirally from the inside to the outside. Here, the disc remains supported only by three small wires.
    (c) Patterning graphene by EBIE with the same writing sequence as in (a). This also leads to a shifted disc (left image). If more dose is applied, disc detaches from the support and folds up (right image, see also supplementary Fig. S3 for intermediate steps).
    Scale bar is (a) \SI{20}{\nano\meter}, (b) \SI{5}{\nano\meter} and (c) \SI{250}{\nano\meter}.}
    \label{fig:quasifree}
\end{figure}

To our surprise, however, it turns out that even structures that contain isolated blocks of material can be written into the TMDs, resulting in a "nearly" free-standing structure held only by very thin, disordered residue of material.  Some examples are shown in Fig.~\ref{fig:quasifree}: Depending on the sequence of writing the pattern, the isolated block of the TMD may remain at its original position, or may shift to one side. However, we have never observed that the isolated block of the TMD materials disappeared entirely. In Fig.~\ref{fig:quasifree}b, the remaining structure is only supported by a few atom thick strings, which appear to be similar to the wires between beam-induced holes of Ref.~\cite{Liu2013a,Lin2014}. Indeed, as noted already in ~\cite{Liu2013a,Lin2014}, these wires are much more resistant to damage from the electron beam than the original TMD lattice. Hence, the process of cutting isolated blocks into TMDs becomes self-limiting at the point where the structure is only hanging on few-atom thick wires of the TMD, and thus can be reproduced reliably. For graphene, in this example structured by EBIE, it is similarly possible to create almost-disconnected pieces of material (Fig.~\ref{fig:quasifree}c). However, in this case, the electron dose must be carefully balanced. Pieces of graphene that are disconnected from the support tend to fold up or adhere next to the hole (supplementary Fig. S3).

\begin{figure}
    \centering
    \includegraphics[width=1.0\textwidth]{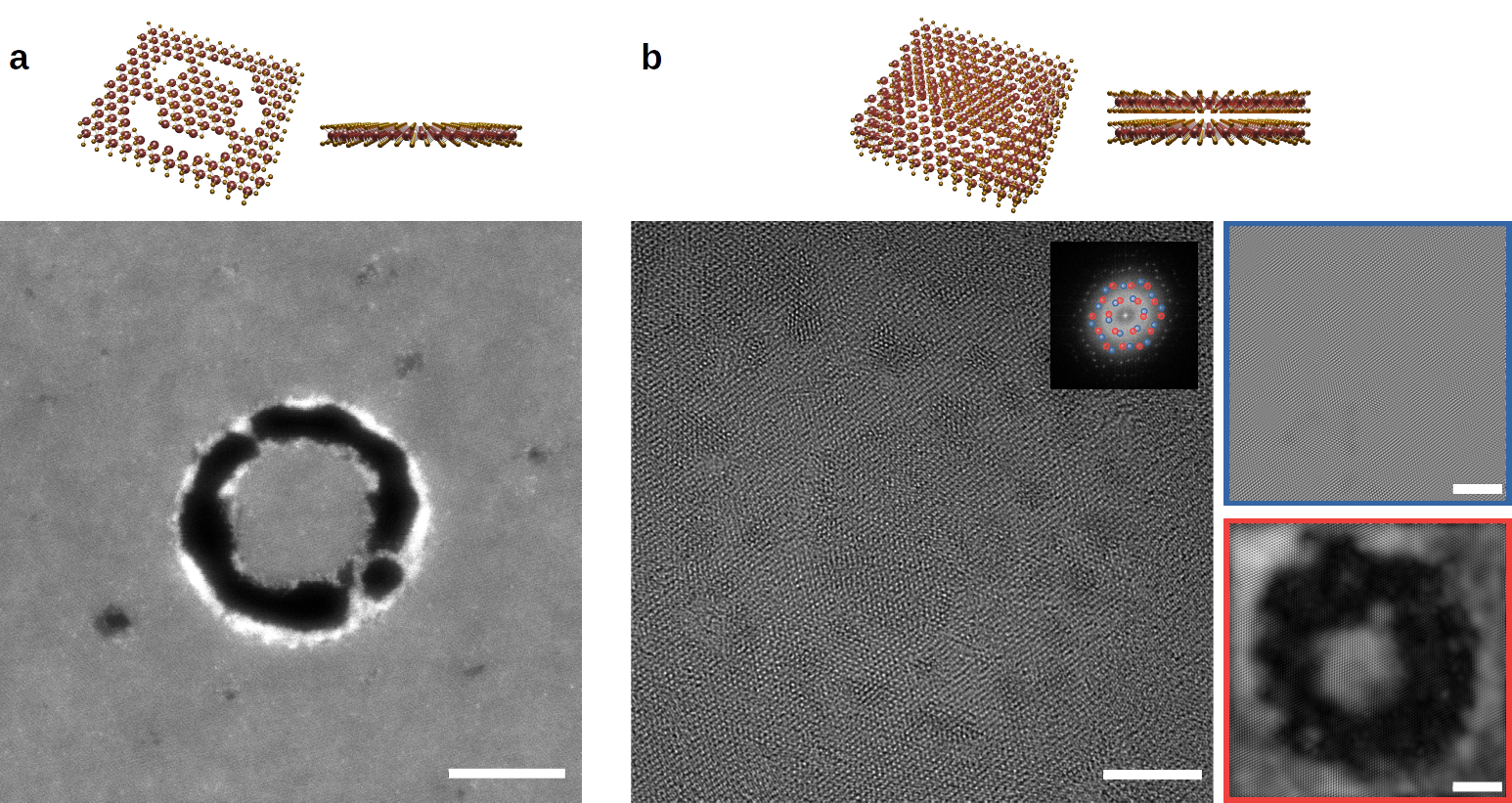}
    \caption{Assembly of non-contiguous structures. (a) A closed ring patterned into a WS$_{2}$ membrane (same as in Fig.~\ref{fig:quasifree}b). Small remaining strings support the inner part and allow the fabrication of almost-free features. (b) Stacking of a non-contiguous WS$_{2}$ structure on a pristine WS$_{2}$ membrane. Fourier filtering of the images is applied to visualize the two individual layers by their different lattice orientations. Scale bar is \SI{5}{\nano\meter}.}
    \label{fig:quasifreeWS2_stacking}
\end{figure}

Even delicate structures as in Fig.~\ref{fig:quasifree}b, where a piece of material is only attached by atomic-scale wires, can be assembled into a stack. This is exemplified in Fig.~\ref{fig:quasifreeWS2_stacking}. After patterning, the almost-isolated WS$_{2}$ quantum dot is stacked on top of another WS$_{2}$ membrane. Fig.~\ref{fig:quasifreeWS2_stacking}b shows the resulting structure. To illustrate the individual layers, Fourier filtering is used to highlight the two different lattice orientations in the atomically resolved image. As shown in Fig.~\ref{fig:quasifreeWS2_stacking}b, one of the layers is continuous, while the other layer contains an isolated disc of WS$_{2}$.

\section{Discussion}

The methods developed in this work enable the fabrication of novel nanostructures and device geometries on the basis of 2D materials. In particular, the aligned stacking of pre-patterned layers extends the existing capabilities for in-plane structuring into the third dimension. Although only three different materials were tested so far, the basic approach can be expected to work with a large variety of 2D materials and possibly other thin layers.

Structuring within each layer can be performed with nearly atomic precision, as evidenced in Fig~\ref{fig:patterning}d, and also in prior works \cite{Fischbein2008,Song2011,Borrnert2012,Xu2013,Liu2014,MasihDas2016,Clark2019a,MasihDas2020}. We also note that the lattice structure of the WS$_{2}$ is practically unaffected outside of the holes that were defined by the focused electron beam. When structuring TMDs by focused, high-energy electron irradiation, we found that one can create isolated blocks of material that are only held in place by a few atomic-scale wires through a self-limiting process. This could be useful to create quantum dot structures out of these materials, or other types of isolated elements in the future.

Stacking of 2D material layers naturally generates an atomic level precision along the third dimension, via the choice of the material, and here also via the potential introduction of gaps in the chosen layers. In this respect, it should be pointed out that in both TEM and SEM imaging of 2D materials, some amount of amorphous carbon is deposited on the 2D materials. This amorphous carbon likely stems from hydrocarbons adsorbed on the sample surfaces which are cracked under electron irradiation. With our experiments carried out in standard TEM and SEM instruments, it is inevitable that these hydrocarbon deposits are also present in the 2D material stacks, including between the layers. Amorphous carbon as formed under an electron beam is not mobile enough to be pushed into pockets of the heterostructure, unlike other contaminants \cite{Haigh2012a,Purdie2018}). Further improvements to avoid this contamination, by carrying out the stacking in an ultra-high vacuum SEM and structuring the sample at elevated temperature, are currently being set up.

The largest remaining uncertainty in the spatial control over the 3D structure is the accuracy in the lateral alignment of the layers. We regularly achieve an alignment accuracy of about 10nm, which is slightly less than what one would expect considering the resolution of the SEM of 2-3nm. The reason is that, currently, the alignment of the layers to each other is only visually estimated by the user while stacking: during the continuous upward movement of the substrate on the Z-stage, the alignment is verified by adjusting the focus to different heights, and corrected using the nanomanipulator if necessary.  In addition, a short pixel time must be selected to observe the process in real time, resulting in rather noisy images. We expect that this could be significantly improved by measuring the offset in the SEM image, e.g. by automated pattern recognition. An additional practical complication when using the SEM stage is a non-perfect decoupling of in-plane (x,y) and height (z) direction, e.g. a jump in (x,y) when inverting the z drive direction. In principle, we see no reason why the accuracy in the lateral alignment couldn't be pushed to the resolution of the observing electron microscope, or even beyond -- indeed, measuring the center position of a given reference feature should be possible with a higher accuracy than the resolution of the underlying image.

The twist angle between features in different layers can also be considered as part of the in-plane alignment. As shown in Fig.~\ref{fig:stackexamples2}g-k, the orientation of the two crosses could be made at the desired relative angle of 45 degrees. This is achieved by patterning the two layers into different locations on the same grid containing the source material (prior to stacking), and then, relative orientation of the features as defined during writing is maintained. Rotation of the target is also possible, but was not used so far.

In this work, we have stacked up to ten layers, with lateral dimensions of the features in the order of 100nm in diameter. This corresponds to a build volume of ca. $3\cdot10^{4}\textrm{nm}^{3}$. With the manual preparation, structuring and placement, the time needed to create this assembly is ca. 8 hours, resulting in a build speed of ca. $1\textrm{nm}^{3}/\textrm{s}$. However, most of this time is used for loading samples into a device, for microscope alignment, and similar steps. If we consider only the actual writing time in the STEM (e.g. 10min for the structure in Fig. \ref{fig:patterning}c), and allow 5 minutes for placing the layer onto the stack, one can estimate a possible speed of ca. $10\textrm{nm}^{3}/\textrm{s}$ for this novel route of 3D structure fabrication. Such a speed could be anticipated if the whole process, including sample transfers, is automated.  This may still appear slow, but it is actually well on the extrapolated trend for 3D printing speed (volume per unit of time) vs. feature size as given in Fig. 7 of Ref. \cite{Liashenko2020}.

\clearpage
\section{Conclusions}

We have presented a new approach to assemble nanopatterned 2D material layers into multi-layer structures. As a key novelty, we have demonstrated that features in subsequent layers can be aligned with a lateral offset of 10nm with the new experimental protocol. The stacking is done in vacuum and under observation of an electron microscope. By the combination of arbitrary structures in each layer and the stacking with precise lateral alignment, it thus becomes possible to create almost arbitrary 3D structures. Moreover, as each layer can be chosen from the large zoo of 2D materials, it should be possible to directly generate a wide variety of functional devices.

\section*{Methods}

\textbf{2D materials transfer on TEM grid.}
Commercial graphene from Graphenea ("easy transfer" mono-layer graphene) is transferred according to the standard protocol of the manufacturer. First, the water soluble polymer is dissolved in deionized water and the floating sacrificial layer with graphene is fished with the TEM grid (QUANTIFOIL® Holey Carbon Films, type R 10/5 - 10µm large holes with 5µm spacing, grid type gold). After drying in air, annealing is performed for one hour at 150°C on a hot plate. The sacrificial layer is removed in acetone (50°C for 1 hour) and isopropyl alcohol (1 hour).

WS$_{2}$ and MoS$_{2}$ are grown by chemical vapor deposition on Si/SiO$_{2}$ Wafer. The TMD flakes are transferred to the TEM grid (same type as above) following the procedure of Ref. \cite{Meyer2008b}. In short, the grid is placed on the wafer, aligned to the flake and held under light pressure with the help of a micromanipulator (MM3A-EM, Kleindiek). A small droplet of isopropyl alcohol is added on top and the surface tension of the solvent during evaporation ensures contact between the flake and the perforated support film. After annealing on a hot plate (200°C, 5 min), the silicon oxide of the wafer is etched away in an aqueous solution of potassium hydroxide. Thus, the TEM grid now carrying the TMD flake detaches from the wafer. The sample is then washed in deionized water and isopropyl alcohol and subsequently dried in air.

\textbf{2D material patterning.}
For patterning the free-standing 2D materials, we use a probe-side aberration corrected JEOL ARM200F or an image-side aberration corrected JEOL ARM200F. Both microscopes are operated in STEM Mode at 200kV for writing the structures. The digital micrograph (DM) software was used for programming and running a script that controls the beam position such that the desired pattern is written into the membrane. After writing the patterns, initial images are obtained at 200kV in STEM mode, while avoiding excessive electron doses. Further images are then acquired at 80kV, using the HRTEM mode on the image-side aberration corrected microscope, or using the STEM mode of the probe-corrected microscope. 
Typical beam currents are 200pA with a typical dose per area of \SI{3e6}{\coulomb\per\meter\squared} for cutting TMDs.

Alternatively, graphene is patterned (after transfer to the TEM grid) by gas-assisted electron beam induced etching in a Zeiss Leo XB1540 as reported by \cite{Sommer2015}. Water vapor is injected near the graphene surface using the gas-injection system of the FIB-SEM (Orsay Physics 5-line GIS). Using an acceleration voltage of 3kV, the electron beam is scanned over the pattern to be created. Oxidation of graphene by ionized species of the etching agent formed by the electron beam lead to volatile COx products which are removed by the vacuum system \cite{Spinney2009}.
For patterning, a beam current of \SI{3}{\nano\ampere} and a typical dose per area of around \SI{5}{\coulomb\per\meter\squared} is used. Since EBIE can be done in the same SEM as the stacking, this simplifies the process, but the resolution is not as good as patterning by STEM.

\textbf{Substrate preparation.}
A commercially available lift-out grid (Omniprobe\textregistered, copper with 5 posts) is used as substrate. The outer posts are bent downwards whereas the center post is bent upwards by about \SI{50}{\micro\meter}. Bending is done by hand using a tweezer and observed in a Zeiss Stemi optical microscope. 

Further preparation of the center post is performed by focused ion beam milling in a Zeiss Crossbeam Auriga 40. The tip of the post is milled narrower to a width of \SI{15}{\micro\meter} and thinner to \SI{5}{\micro\meter} using an acceleration voltage of \SI{30}{\kilo\volt} and a milling current of \SI{10}{\nano\ampere}. With a lower current of \SI{500}{\pico\ampere}, a \SI{2.5}{\micro\meter} x \SI{2.5}{\micro\meter} hole is milled as viewing window in TEM (see also supplement Fig. S1).

\textbf{Stack assembly.}
The layer assembly is performed in a Zeiss Leo XB1540 equipped with a Kleindiek MM3A-EM nanomanipulator. The TEM grid carrying the 2D material to be stacked is clamped by the manipulator, which is mounted horizontally on the SEM chamber. The lift-out grid as substrate for stacking is held by the SEM six-axis stage. Before loading in the SEM for the assembly, the samples are baked out on a hot plate at 200°C for 10 minutes.
The SEM is operated at an acceleration voltage of 15kV to enable the observation of both the membrane to be deposited as well as the target membranes. For the alignment, the Inlens-SE detector is mixed with the STEM detector and a dual-magnification mode is used that alternately records a normal image and then an image with a smaller field of view (see video in supplement). This enables to observe both the whole support frame as well as introduced feature in the membranes in more detail simultaneously.

\section*{Acknowledgments}

We acknowledge funding by the Ministry of Science, Research and
Art Baden-Württemberg, by the State of Baden-Württemberg, Germany
and by the European Union under EU-EFRE grant no. 712889.
We would like to thank Birgit Schröppel and Claus J. Burkhardt of the NMI for their support on the scanning electron microscopes. 

\section*{Supporting information}
Preparation of the target substrate, SEM images of ten layer graphene assembly, SEM images of patterning "nearly" free-standing discs by EBIE, Video of stacking two layers in SEM.

\bibliography{library}

\end{document}


\section*{Supporting information}

\subsection*{Preparation of the target substrate}
\begin{figure}[b!]
\includegraphics[width=0.8\textwidth]{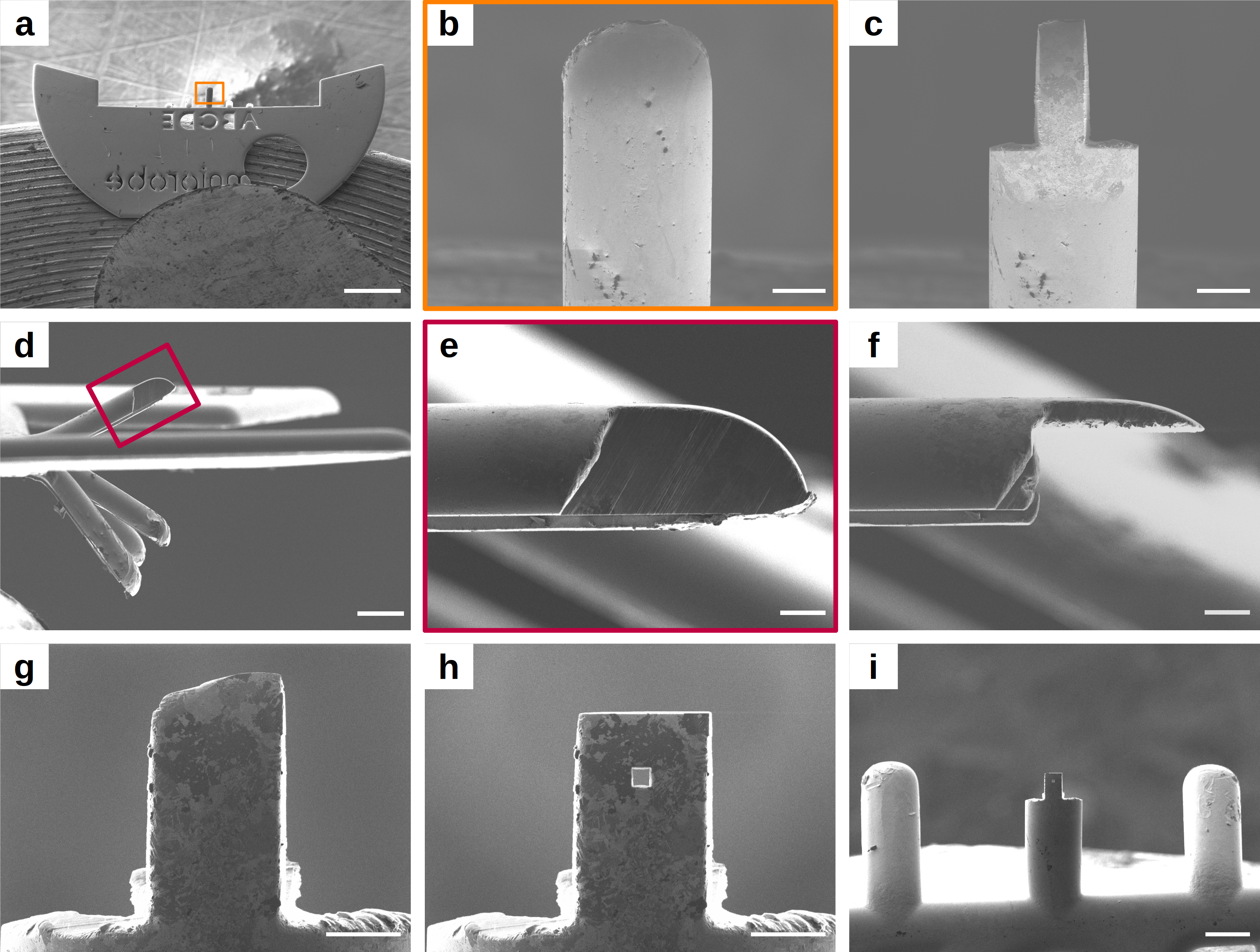}
\centering
\caption{
Preparation of the target substrate. (a) Lift-out grid with one of the five post bent upwards, while the others are bent downwards (not visible). (b) Original shape of the post. (c) Post cut to a smaller width by FIB. (d) Side view of the lift-out grid. Central post (e) before and (f) after milling from the side to reduce the thickness. Top view of the post (g) before and (h+i) after cutting the central hole for material deposition.
Scale bars are (a) \SI{500}{\micro\meter}, (b+c) \SI{20}{\micro\meter}, (d+i) \SI{50}{\micro\meter}, (e+f+g+h) \SI{10}{\micro\meter}. (a+i) are SEM images (SE-detector), (b-h) are FIB images (30kV, 50pA).
}
\label{fig:substrateprep}
\end{figure}

The preparation of the target substrate is illustrated in Fig. \ref{fig:substrateprep}. A single post in a FIB lift-out grid is thinned and narrowed by FIB, and a hole is inserted. It is important that this hole is made on the curved surface near the end of the post. 

\clearpage
\subsection*{Example of the layer-by-layer assembly}
\begin{figure}[h!]
    \centering
    \includegraphics[width = 1\linewidth]{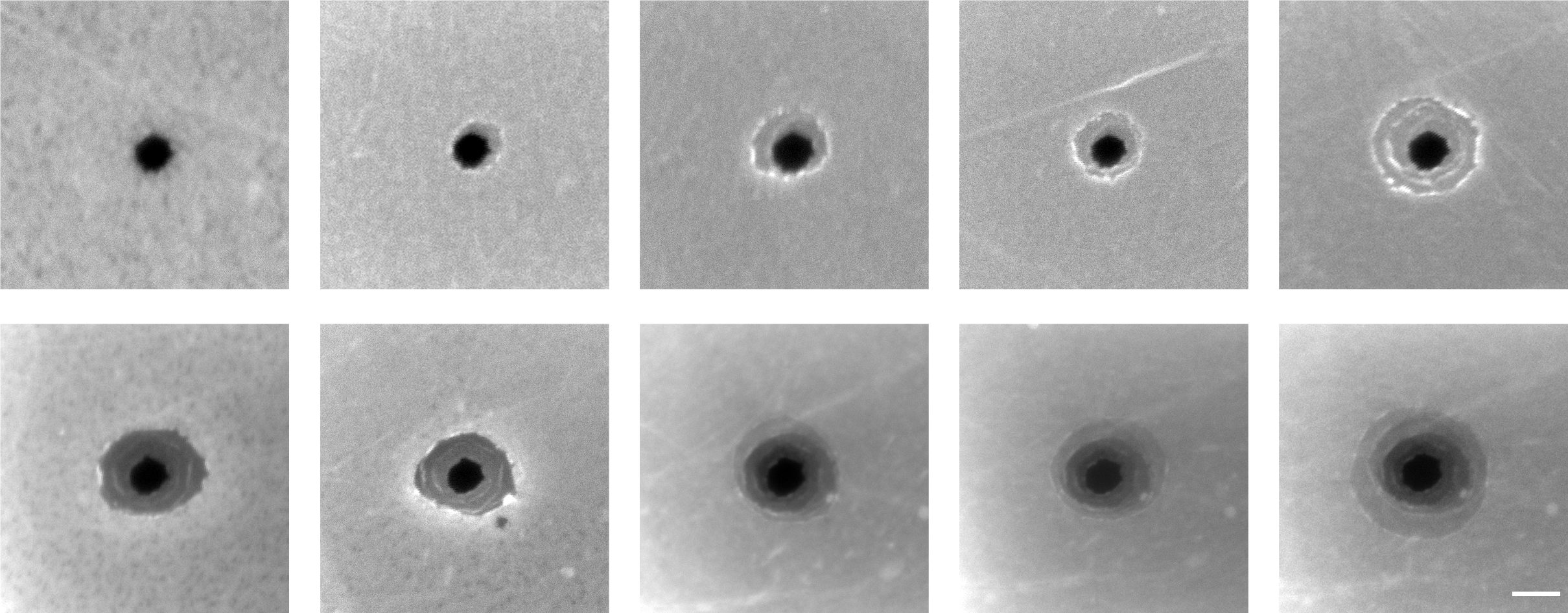}
    \caption{Layer-by-layer assembly of ten graphene layers with patterned holes of increasing diameter. After placing each layer, a SEM image is taken. The offset, i.e. the distance between the centers of the holes, is determined for each layer with respect to the first layer (smallest hole). Starting from the second layer to the last layer it is 19, 18, 15, 14, 16, 7, 23, 35, 7nm resulting in an average offset for the ten layers of about 17nm.
    Scale is \SI{200}{\nano\meter}.}
     \label{fig:tenlayers}
\end{figure}

For the presented nanoscale funnel, ten graphene layers are patterned by EBIE with holes of different sizes and subsequently placed on top of each other in the order of increasing hole size. The stepwise assembly is shown in Fig. \ref{fig:tenlayers} as SEM image sequence after each assembled layer. 

\clearpage
\subsection*{Patterning non-contiguous structures}
\begin{figure}[h!]
    \centering
    \includegraphics[width = 1\linewidth]{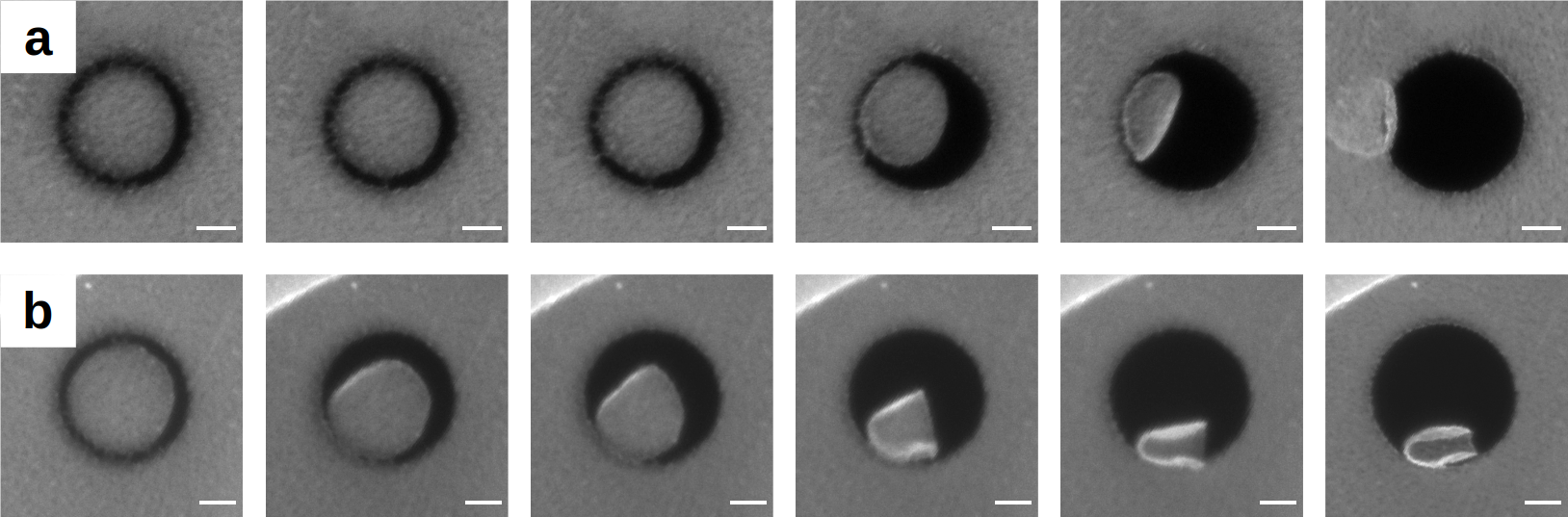}
    \caption{Patterning "nearly" free-standing discs by EBIE. After an almost isolated structure is formed, additional electron dose causes the structure to (b) adhere next to the hole or (a) to fold up.
    Scale is \SI{250}{\nano\meter}.}
     \label{fig:circlesEBIE}
\end{figure}

The patterning of a closed ring into graphene by means of EBIE results in an almost free-standing disc, as shown in each first image of Fig. \ref{fig:circlesEBIE}a and b. The SEM image sequence shows the effect of additional electron dose. In Fig. \ref{fig:circlesEBIE}a, the cut-out structure first folds up and then adheres next to the hole on the graphene membrane, whereas in Fig. \ref{fig:circlesEBIE}b the folded structure remains curled at the edge of the hole.

